\documentclass[5p]{elsarticle}

\usepackage{hyperref}
\usepackage{amsmath}
\usepackage{color}
\usepackage{xfrac}
\usepackage[T1]{fontenc}
\usepackage{upgreek}
\newcommand{\mus}         {{$\upmu$s}}

\newcommand{\qbb}         {{$Q_{\beta\beta}$}}

\newcommand{\onbb}        {{$0\nu\beta\beta$}}

\newcommand{\gerda}       {\textsc{Gerda}}
\newcommand{\infn}       {\textsc{Infn}}

\newcommand{\lngs}        {{\mbox{\textsc{Lngs}}}}
\newcommand{\cngs}        {{\mbox{\textsc{Cngs}}}}

\newcommand{\WT}          {water tank}
\newcommand{\borexino}    {\mbox{\textsc{Borexino}}}

\journal{Astroparticle Physics}
\begin{document}

\begin{frontmatter}

\title{Flux Modulations seen by the Muon Veto of the \gerda\  Experiment}

%  authors per affiliation:
\author{\mbox{\protect\textsc{Gerda}}
  collaboration\fnref{aalab}
\corref{mycorrespondingauthor}%\fnref{aalab}
\cortext[mycorrespondingauthor]{Correspondence: {gerda-eb@mpi-hd.mpg.de}}}
\fntext[aalab]{Laboratori Nazionali del Gran Sasso, Assergi, Italy}
%\ead{gerda-eb@mpi-hd.mpg.de}
%
\author[addtum]{~\\[3mm]M.~Agostini}
\author[adddre]{M.~Allardt}
\author[addku]{A.M.~Bakalyarov}
\author[addlngs]{M.~Balata}
\author[addinr]{I.~Barabanov}
\author[adddre]{N.~Barros\fnref{nowPEN}}
\author[adduzh]{L.~Baudis}
\author[addhd]{C.~Bauer}
\author[addmpp]{N.~Becerici-Schmidt}
\author[addmibf,addmibinfn]{E.~Bellotti}
\author[additep,addinr]{S.~Belogurov}
\author[addku]{S.T.~Belyaev}
\author[adduzh]{G.~Benato}
\author[addpduni,addpdinfn]{A.~Bettini}
\author[addinr]{L.~Bezrukov}
\author[addtum]{T.~Bode}
\author[addcr,addjinr]{D.~Borowicz}
\author[addjinr]{V.~Brudanin}
\author[addpduni,addpdinfn]{R.~Brugnera}
\author[addmpp]{A.~Caldwell}
\author[addminfn]{C.~Cattadori}
\author[additep]{A.~Chernogorov}
\author[addlngs]{V.~D'Andrea}
\author[additep]{E.V.~Demidova}
\author[addlngs]{A.~di~Vacri}
\author[adddre]{A.~Domula}
\author[addinr]{E.~Doroshkevich}
\author[addjinr]{V.~Egorov}
\author[addtue]{R.~Falkenstein}
\author[addinr]{O.~Fedorova}
\author[addtue]{K.~Freund}\fnref{nowMaxment}
% ...
\address[addlngs]{INFN Laboratori Nazionali del Gran Sasso, LNGS, and Gran
  Sasso Science Institute, GSSI, Assergi, Italy}
\address[addals]{INFN Laboratori Nazionali del Sud, Catania, Italy}
\address[addcra]{Institute of Physics, Jagiellonian University, Cracow, Poland}
\address[adddre]{Institut f{\"u}r Kern- und Teilchenphysik, Technische
  Universit{\"a}t Dresden, Dresden, Germany}
\address[addjinr]{Joint Institute for Nuclear Research, Dubna, Russia}
\address[addgeel]{Institute for Reference Materials and Measurements, Geel,
     Belgium}
\address[addhd]{Max-Planck-Institut f{\"u}r Kernphysik, Heidelberg, Germany}
\address[addmibf]{Dipartimento di Fisica, Universit{\`a} Milano Bicocca,
     Milan, Italy}
\address[addmibinfn]{INFN Milano Bicocca, Milan, Italy}
\address[adduminfn]{Dipartimento di Fisica, Universit{\`a} degli Studi di
  Milano e INFN Milano,  Milan, Italy}
\address[addinr]{Institute for Nuclear Research of the Russian Academy of
  Sciences, Moscow, Russia}
\address[additep]{Institute for Theoretical and Experimental Physics,
    Moscow, Russia}
\address[addku]{National Research Centre ``Kurchatov Institute'', Moscow,
  Russia}
\address[addmpp]{Max-Planck-Institut f{\"ur} Physik, Munich, Germany}
\address[addtum]{Physik Department and Excellence Cluster Universe,
    Technische  Universit{\"a}t M{\"u}nchen, Munich, Germany}
\address[addpduni]{Dipartimento di Fisica e Astronomia dell{`}Universit{\`a}
  di Padova, Padova, Italy}
\address[addpdinfn]{INFN  Padova, Padova, Italy}
\address[addtue]{Physikalisches Institut, Eberhard-Karls Universit{\"a}t
  T{\"u}bingen, Germany}
\address[adduzh]{Physik Institut der Universit{\"a}t Z{\"u}rich, Z{\"u}rich,
    Switzerland}
\fntext[nowPEN]{\emph{present address:} Dept. of Physics and Astronomy,
  U. of Pennsylvania, Philadelphia, Pennsylvania, USA}
\begin{abstract}
The \gerda\ experiment at \lngs\ of INFN is equipped with an active muon veto.
The main part of the system is a water Cherenkov veto with 66~PMTs in the
\WT\ surrounding the \gerda\ cryostat. The muon flux recorded by this veto
shows a seasonal modulation. Two effects have been identified which are caused
by secondary muons from the \cngs\ neutrino beam (2.2\,\%) and a temperature
modulation of the atmosphere (1.4\,\%).  A mean cosmic muon rate of
$I^0_{\upmu}=(3.477\pm0.002_{\textrm{stat}}\pm
0.067_{\textrm{sys}})\times10^{-4}$/(s$\cdot$m$^{2}$) was found in good
agreement with other experiments at \lngs\ at a depth of 3500~meter water
equivalent.
\end{abstract}

\begin{keyword}
water Cherenkov detector \sep underground experiment \sep cosmic rays \sep
muon interaction \sep 
\end{keyword}
\end{frontmatter}

%%%%%%%%%%%%%%%%%%%%%%%%%%%%%%%%%%%%%%%%%%%%%%%%%%%%%%%%%%%%%%%%%%%%%%%%%%%%%%%%
\section{Introduction}
      \label{sec:intro}
The \gerda\ (\textbf{Ger}manium \textbf{D}etector \textbf{A}rray) experiment
is searching for the neutrinoless double-beta (\onbb) decay of
$^{76}$Ge~\cite{gerda_tec,gerda_prl}. It is located in Hall A of the
underground laboratory Laboratori Nazionali del Gran Sasso (\lngs) of
\infn\ at a depth of 3500 meter water equivalent (m.w.e.). In order to search
effectively for such a rare process as the \onbb\ decay \gerda\ needs to be
equipped with a dedicated veto systems which tags muons passing the
experiment~\cite{gerda_tec,muon_vetotec,kai}.  Here we report about the
observed rates including an annual modulation of the latter during the period
2010--2013 which encompasses Phase~I of \gerda. Other underground experiments
have observed similar annual modulations of their rates, either due to the
muon flux~\cite{macro_mu_93} or of other origin~\cite{dama}.

A full explanation of the muon rate is important to assure that the
systematics of the experiment are fully understood, in particular when aiming
for reduced backgrounds in future phases.  A particular problem could be the
generation of unstable isotopes by the muons directly or through the secondary
neutron flux. Thus, the more obvious sources of backgrounds must be
understood. The present results serve also as cross checks to previous or
future data sets on muon fluxes in underground laboratories.

\section{Modulations}
 \label{sec:modu}

The hardware of the muon veto worked very reliable and stable.  The overall
muon rate of the veto is observed to be modulated by two different sources.
Firstly, the majority of the detectable muons are produced
cosmogenically~\cite{gaisser}. Their spectrum and angular distribution within
the halls are both altered by the profile of the rock overburden and have been
measured for \lngs\ with high precision~\cite{macro_mu_93}. These muons
have an average energy of $\langle{E_\mu}\rangle=270$~GeV. Due to seasonal
temperature changes in the atmosphere the mean muon energy changes over the
year and thus the muon flux at \lngs.

% For the interpretation this analysis relies on a model by

Secondly, an artificial source for muons was the \textsc{Cern} Neutrinos to
Gran Sasso (\cngs) neutrino beam~\cite{cngs13} in the period 2008--2012
serving the \textsc{Opera} experiment~\cite{opera} for the search of
$\nu_\mu\rightarrow\nu_\tau$ oscillations. This $\nu_\mu$ beam can create
muons by charged current reactions along its 730~km long path. Thus, an
additional muon flux is expected during any \cngs\ beam line operation. As the
beam is not operated continuously but pauses in the winter months, the
additional flux takes the form of an annual modulation.

Both effects can be described with high precision. The parameters for the
atmospheric muon generation are presented in this work which agree well with
other experiments at \lngs.
%%%%%%%%%%%%%%%%%%%%%%%%%%%%%%%%%%%%%%%%%%%%%%%%%%%%%%%%%%%%%%%%%%%%%%%%%%%%%%%%
\section{Instrumentation}
 \label{sec:instru}
\begin{figure}[t]
\begin{center}
  \includegraphics[width=.7\columnwidth]{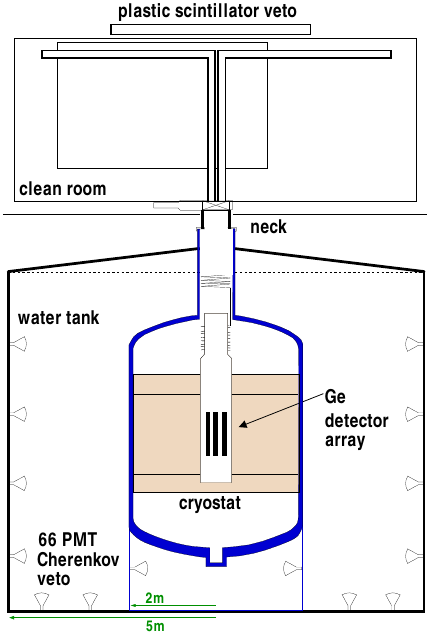}
  \caption{   \label{fig:exp}
         Sketch of the \gerda\ experiment~\cite{gerda_tec,muon_vetotec}.
}
\end{center}
\end{figure}

The muon veto consists of two parts.  The \gerda\ \WT\ is instrumented with 66
photomultipliers of 8'' size which detect the Cherenkov light of passing
muons~\cite{gerda_tec,muon_vetotec} (Fig.~\ref{fig:exp}). The germanium
crystals are lowered through the ``neck'' from the clean room into the liquid
argon (LAr) cryostat for normal operation.  This ``neck'' region is less
monitored by the Cherenkov veto. Thus, it is covered in addition by a
4$\times$3~m$^2$ layer of plastic scintillators equipped with PMTs on top of
the \gerda\ clean room. In the analysis shown here the standard \gerda\ muon
veto analysis cuts were performed. As described in Ref.~\cite{muon_vetotec} a
18~p.e. (photo electron) cut on the Cherenkov data or a valid combination of
panels are needed to create a trigger. The stability of the rates and of the
light output of veto system was checked periodically.

This muon veto data set contains a period of 806 days from November 2010 to
July 2013 that includes also a period before Phase~I. Particularly during
Phase~I of the \gerda\ experiment the veto system ran continuously stable and
reliable.

%%%%%%%%%%%%%%%%%%%%%%%%%%%%%%%%%%%%%%%%%%%%%%%%%%%%%%%%%%%%%%%%%%%%%%%%%%%%%%%%
\begin{figure}[t]
\begin{center}
  \includegraphics[width=.46\textwidth]{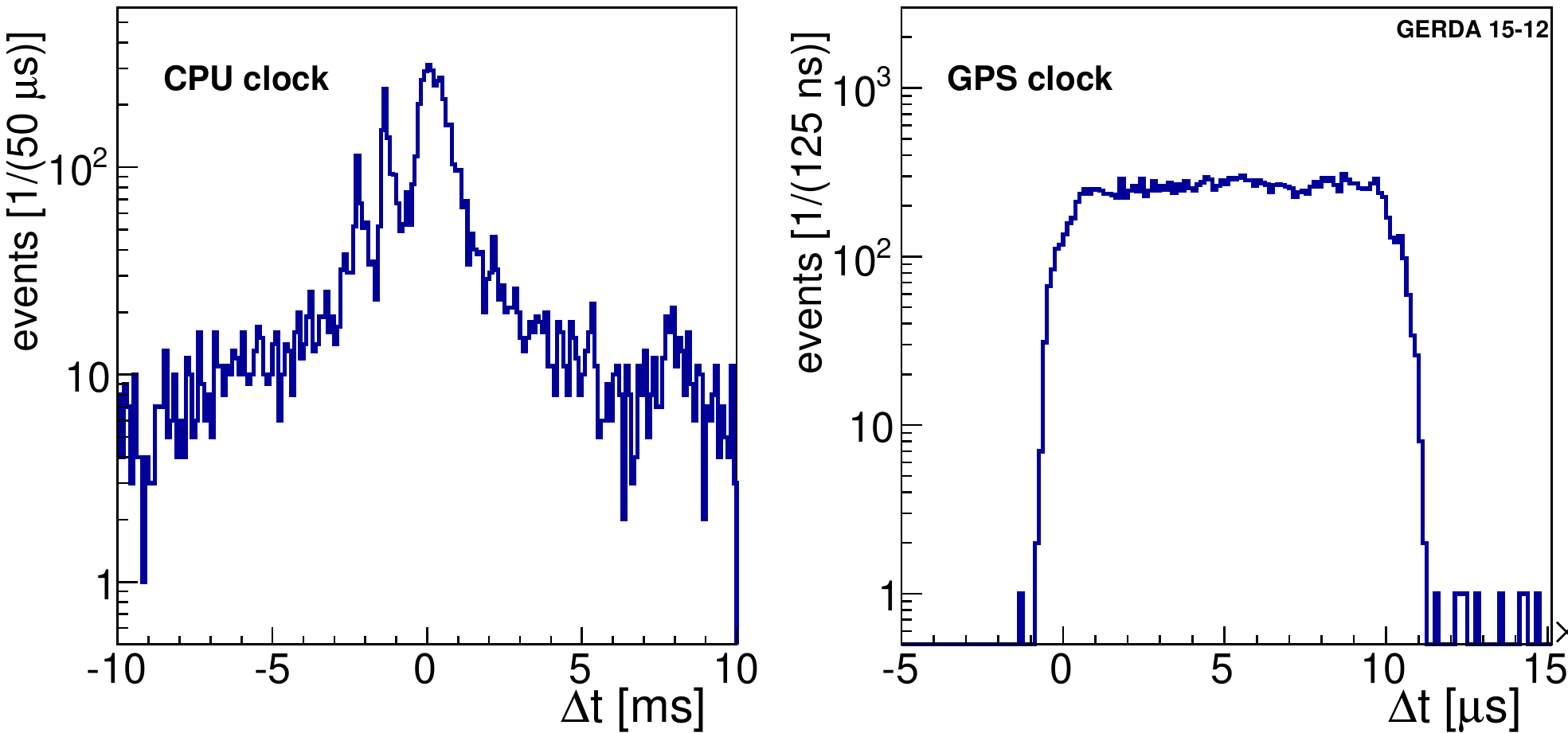}
  \caption{   \label{fig:timeres}
         Time offset between \textsc{Cngs} and muon veto events. The left plot
         shows the time resolution before the installation of the GPS clock,
         the right one after. The origin in both plots is set to the rising
         flank of the main feature. Note the different time scales.
}
\end{center}
\end{figure}
%%%%%%%%%%%%%%%%%%%%%%%%%%%%%%%%%%%%%%%%%%%%%%%%%%%%%%%%%%%%%%%%%%%%%%%%%%%%%%%%

%%%%%%%%%%%%%%%%%%%%%%%%%%%%%%%%%%%%%%%%%%%%%%%%%%%%%%%%%%%%%%%%%%%%%%%%%%%%%%%%
\begin{figure*}[tH!]
\begin{center}
  \includegraphics[width=\textwidth]{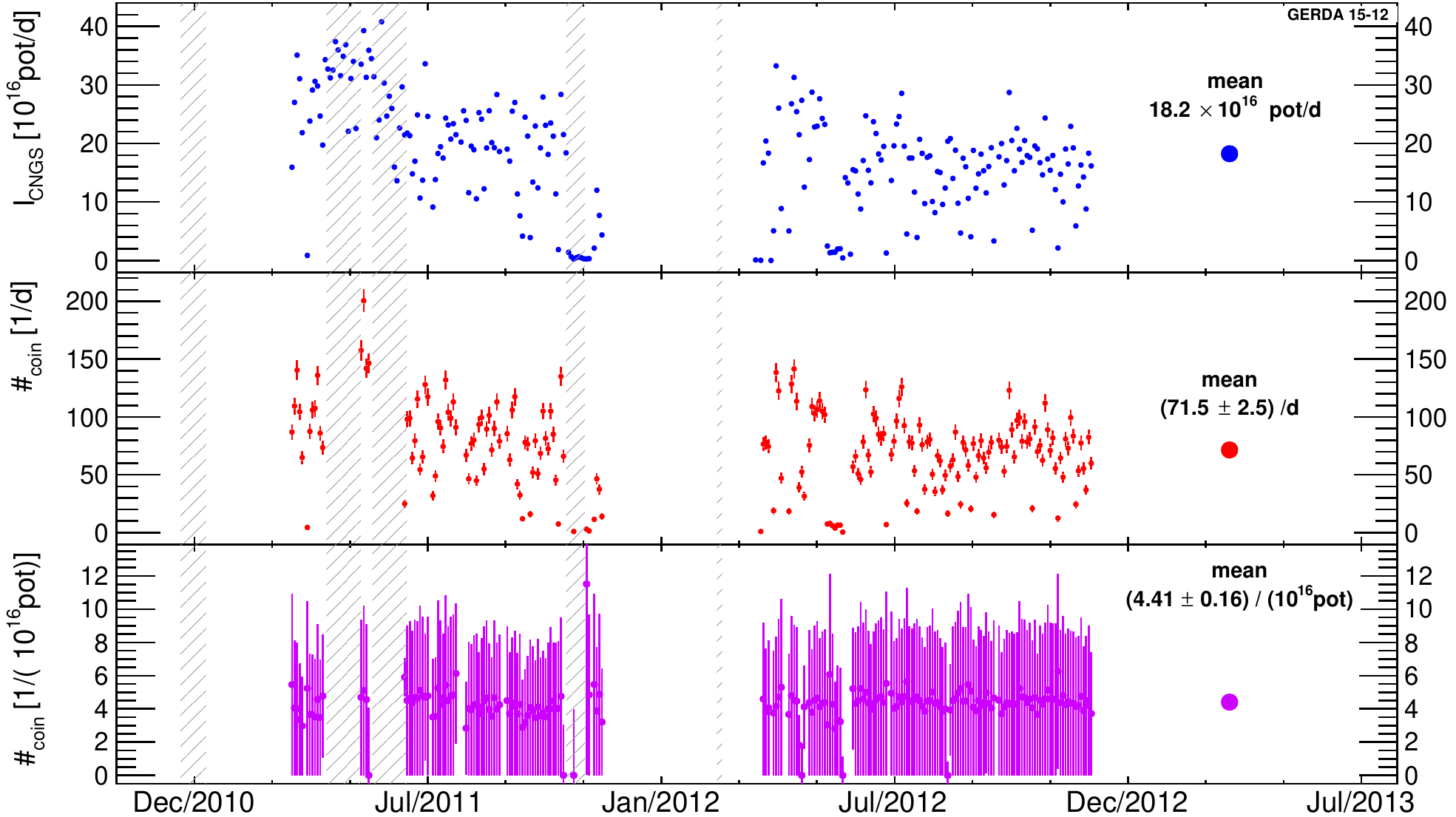}
  \caption{    \label{fig:cngsint}
         \textsc{Cngs} beam intensities in protons-on-target (POT) and rates
         over time with a binning of two days. Top: beam intensities measured
         at \textsc{Cern}; middle: events correlated with the muon veto; bottom:
         ratio of the two. The grey hatched areas indicate breaks in the muon
         data-taking.
}
\end{center}
\end{figure*}
%%%%%%%%%%%%%%%%%%%%%%%%%%%%%%%%%%%%%%%%%%%%%%%%%%%%%%%%%%%%%%%%%%%%%%%%%%%%%%%%

%%%%%%%%%%%%%%%%%%%%%%%%%%%%%%%%%%%%%%%%%%%%%%%%%%%%%%%%%%%%%%%%%%%%%%%%%%%%%%%%
\section{Influence of the CNGS beam}

The \textsc{Cern} SPS delivered proton bunches with an energy of 400~GeV that
hit a carbon target in the \cngs\ beam line~\cite{cngs13}. Actually, each SPS
extraction consists of two proton bunches which are 10.5~\mus\ wide and 50~ms
apart. Normally, the extraction is repeated every 6~s. Pions and kaons from
the collision products are focused on a decay line pointed towards
\lngs. These particles can decay according to $\pi^+/K^+ \rightarrow \mu^+ +
\nu_\mu$. Muon detectors at the end of the decay line record the $\mu^+$ which
can be correlated with the $\nu_\mu$ intensity. The primary $\mu^+$ will be
stopped on the way towards \lngs\ while the $\nu_\mu$ travel almost
unhindered. The $\nu_\mu$ however are able to produce secondary muons of
$\langle{E_\mu}\rangle=17$~GeV via $\nu_\mu + d \rightarrow \mu^- + u$
reactions upstream  of \lngs. Thus, an additional muon flux
impinges horizontally into the \gerda\ setup.

Both systems, the \cngs\ beam at \textsc{Cern} and the muon veto of the
\gerda\ experiment, were operational at the same time during 404 days
in 2011 and 2012. In this period $\sim$28800 coincident muons due to
$\sim$$7\times10^{19}$ protons-on-target were detected. The events of both
\cngs\ and muon veto were correlated by using their respective time stamps. In
Fig.~\ref{fig:timeres} the time differences of valid signals from both
facilities are shown. The enhancement above the random background shows that 
true coincidences are observed. In the beginning, the muon veto was running
with a \textsc{Cpu} clock. However, prior to the start of Phase~I a
\textsc{Gps} clock was installed~\cite{gerda_tec}. A sharpening of the
enhancement of the correlated events is clearly visible from the time spectra
of both clock systems.  Compared to the \textsc{Cpu} clock
(Fig.~\ref{fig:timeres}, left) the time resolution increased dramatically
after the installation of the \textsc{Gps} clock (Fig.~\ref{fig:timeres},
right). With the \textsc{Gps} clock the 10.5~\mus\ bunch length of the
\cngs\ beam can be reproduced. This shows that the recorded events can be
correlated in time with high accuracy when tested against an external source
like the \cngs\ beam. The accurate timing will also be of advantage when
searching for cosmogenic reaction products and their identification via their
half life.

In Fig.~\ref{fig:cngsint} time series of the daily beam intensities measured
at \textsc{Cern} (top), the number of events correlated with the muon veto per
day (middle) and the number of coincident events per beam intensity (bottom)
are shown. The flat distribution in the bottom panel demonstrates the
proportionality between the beam intensity and muon
%%%%%%%%%%%%%%%%%%%%%%%%%%%%%%%%%%%%%%%%%%%%%%%%%%%%%%%%%%%%%%%%%%%%%%%%%%%%%%%%
\begin{figure*}[t]
  \begin{center}
  \includegraphics[width=\textwidth]{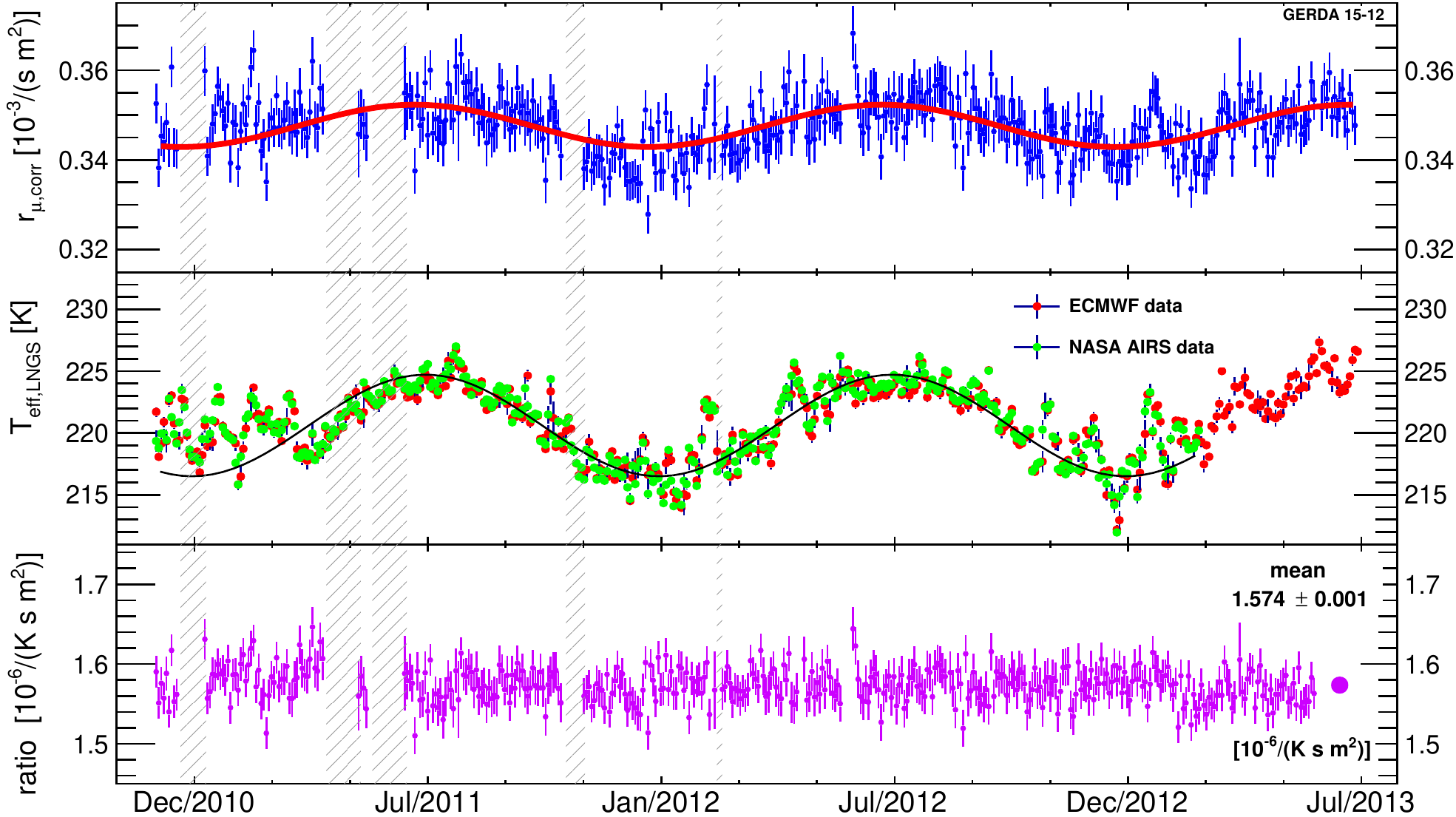}
  \caption{   \label{fig:modulation}
       Top: Muon flux measured by \gerda\ with a binning of two days
       corrected for the \textsc{Cngs} events. A cosine with a period
       of 365.25 days is fit to the data. % ($\chi^2$/ndf=522/408).
       Middle: The effective temperature $T_{\textrm{eff}}$ for muon
       production derived  from data of \textsc{Ecwmf}~\cite{ecwmf} in red
       and from \textsc{Airs}~\cite{airsweb} in green. The black line is a fit
       to the \textsc{Airs} data.
       Bottom: the ratio of the muon rate and the $T_{\textrm{eff}}$ from the
       \textsc{Ecwmf} data set is shown to be flat over the entire time.
}
  \end{center}
\end{figure*}
%%%%%%%%%%%%%%%%%%%%%%%%%%%%%%%%%%%%%%%%%%%%%%%%%%%%%%%%%%%%%%%%%%%%%%%%%%%%%%%%
%\noindent
  events.
This nicely confirms the correct identification of coincident  events. The
overall fraction of \cngs\ events in the muon data is of the order of 2.2\,\%.
%\clearpage

Using the time stamps, coincidences between germanium data and \cngs\ events
can be found as well. A number of 45 coincident germanium events were
identified in Phase~I of \gerda\ and 42 of these were accompanied by a muon
veto trigger and hence correctly discarded. With the rates of the beam and the
germanium detectors a number of $4.9\pm2.2$ random coincident events are
expected for this period and thus the remaining three germanium events can be
attributed to random coincidences.

A single \cngs\ event was recorded in the 230~keV wide interpolation region
around \qbb\ for the background-index (BI) of \gerda. However, this event had
a veto flag and was hence excluded from the analysis and had no effect on the
BI. Since the \cngs\ has been decommissioned in 2013, there will be no
influence of this type in the next phase of \gerda.
 
%%%%%%%%%%%%%%%%%%%%%%%%%%%%%%%%%%%%%%%%%%%%%%%%%%%%%%%%%%%%%%%%%%%%%%%%%%%%%%%%
\section{Atmospheric temperature  modulation}

The identified events due to the \cngs\ beam shown in Fig.~\ref{fig:cngsint}
are removed from the sample for a proper analysis of the temperature
dependence of the cosmic muon fraction (see Fig.~\ref{fig:modulation}, top).
The annual modulation of the muon flux is a well-studied
phenomenon~\cite{gaisser,grashorn,minos_var,borex_modu}. Due to the shielding
effect of atmosphere and rock overburden, only cosmogenically produced muons
with an energy above a certain threshold $E_{\textrm{thr}}$ will be able to
reach an experiment located at a specific depth. Most muons are decay products
of pions and kaons in showers caused by cosmic radiation. The amount of energy
that can be transferred to their decay products depends on the number of
scatterings of the mesons during their life time. The number of scatterings of
a meson is governed by its mean free path which depends on the density of the
air and ultimately is influenced by the temperature. Hence the atmospheric
temperature and the subsequent density of air molecules influence the muon
energy spectrum and thus the flux at a certain depth. Since the main
temperature change is seasonal, in first order approximation a cosine-like
behavior of the flux can be assumed that takes the form
\begin{equation}
  I_{\upmu}(t)=I^0_{\upmu}+\delta I_{\upmu} \cos
  \left(\frac{2\pi}{T}(t-t_0)\right), %I^0_{\upmu}+\Delta I_{\upmu}=
\end{equation}%
where $I_{\upmu}(t)$ is the actual,  $I^0_{\upmu}$ the mean muon flux, and
$\delta I_{\upmu}$ the amplitude of the modulation; $t_0$ is the phase marking
the summer maximum.

The fit to the rate of the \gerda\ muon veto is shown in
Fig.~\ref{fig:modulation} (top).  The period of the fit was fixed to
$T=365.25$~d because only two maxima are covered up to now.  From the top
panel it is obvious that a pure cosine-function will not describe the rate
modulation due to local weather conditions changing from year to year, like an
unusually warm winter 2010/11. The same kind of deviations can be observed
also from the temperature data (Fig~\ref{fig:modulation}, middle).

Muon fluxes at  \lngs\ are conventionally given normalized to the effective
area of the experiment, i.e. the projection of its geometry on the muon
angular spectrum, weighted by the muon intensity. The \gerda\ \WT\ has an
effective area of $(103.5\pm2.0)$~m$^2$. The fit yields a muon flux of
$I^0_{\upmu}=(3.477\pm0.002_{\textrm{stat}}\pm
0.067_{\textrm{sys}})\times10^{-4}$/(s$\cdot$m$^{2}$). The systematic error is
derived solely from the uncertainty of the effective area of the \WT. Due to a
muon detection efficiency of nearly unity~\cite{muon_vetotec} its
contributions to the systematic uncertainty becomes negligible.  The
modulation of $\delta I_{\upmu}=(4.72\pm0.33)\times10^{-6}$/s corresponds to
an amplitude of $(1.4\pm0.1_{\textrm{stat}})\,\%$ of $I^0_{\upmu}$. The phase
$t_0$ shifts the maximum to the 10$^{\textrm{th}}$ of July ($\pm4$~days).  The
fit parameters for the muons agree well with results from other experiments at
\lngs, some of them are listed in Tab.~\ref{tab:modupar}. The phase of all
experiments has its maximum in early July and hence it is not compatible with
the phase observed by the \textsc{Dama} dark matter experiment that is on the
2$^{\textrm{nd}}$ of June~\cite{dama}.
 
The deviation $\Delta I_{\upmu}(t)=I_{\upmu}(t)-I^0_{\upmu}$ from the detected
average muon flux depends on the change in temperature $\Delta T(X,t)$ of a
given layer $X$ of the atmosphere. The overall change of the muon flux can
then be written as an integral over all layers:
\begin{equation}\label{eq:sec}
  \Delta I_{\upmu}(t)=\int_0^{\infty}dX\,\, W(X)\,\, \Delta T(X,t) 
\end{equation}
The coefficient $W(X)$ (see Ref.~\cite{grashorn} for details) contains both
the weight of a certain atmospheric layer to the overall muon flux for both
pions and kaons as well as the threshold energy given for a certain
underground site, i.e. the rock overburden. The effective temperature
$T_{\textrm{eff}}$ is a weighted average of the temperature over all layers of
the atmosphere assuming  the atmosphere to be an isothermal
body~\cite{grashorn}. It can be approximated as follows:
\begin{equation}\label{eq:third}
T_{\textrm{eff}}(t) = \frac{\int_0^{\infty}dX\,\, W(X)\,\, 
  T(X,t)}{\int_0^{\infty}dX\,\, W(X)}
\end{equation}%
Similarly for the difference:% $\Delta T_{\textrm{eff}}(t)$:
\begin{equation}\label{eq:four}
\Delta T_{\textrm{eff}}(t) = \frac{\int_0^{\infty}dX\,\, W(X)\,\, \Delta 
  T(X,t)}{\int_0^{\infty}dX\,\, W(X)}
\end{equation}%
There are two sets of temperature data available for the period studied. The
European Centre for Medium-range Weather Forecast
(\textsc{Ecwmf})~\cite{ecwmf} offers climate data taken by many different
observational methods such as weather stations, aircrafts, balloons and
satellites to interpolate the climate at any given location. For each
location, temperature data for 37 atmospheric pressure levels from 0--1000~hPa
are listed four times a day. The second data set is provided by the
\textsc{Airs} instrument~\cite{airsweb} on-board the \textsc{Nasa Aqua}
satellite~\cite{airs}. The analyses regarding the \textsc{Airs} data were
produced with data retrieved with the \textsc{Giovanni} online data system,
developed and maintained by the \textsc{Nasa Ges Disc}~\cite{giovanni}. The
satellite is in a synchronous orbit with the sun and thus it passes each
position of the earth twice per day. The ascending overpass over the Gran
Sasso is at about 1:00 a.m. and the descending overpass at 1:00
p.m. \textsc{Airs} is an infrared sounder and can therefore be disturbed by
clouds. Similar to \textsc{Ecwmf} it provides temperature data in 24 different
pressure levels at any given point. The effective temperature calculated from
both datasets can be seen in the middle panel of
Fig.~\ref{fig:modulation}. Both sets agree very well with their trends and
also within their fine-structure despite their different detection and
analysis methods.  A fit of a cosine-function yields a mean temperature of
$T_\mathrm{eff}^{0,\textsc{Airs}}=220.6\pm0.2$~K for the \textsc{Airs} data
and $T_\mathrm{eff}^{0,\textsc{Ecwmf}}=221\pm1$~K for the \textsc{Ecwmf}
data. Both amplitudes are found to be 4~K and the temperature maxima are found
on day $t_0=186\pm0.5$, i.e. July 4$^{\textrm{th}}$. Given the short period
and the gaps, the agreement between the maxima for the muon rate and for the
temperature is more than adequate.

The bottom panel of Fig.~\ref{fig:modulation} gives the ratio of the muon rate
and the effective temperature. This ratio is constant for the measured period
yielding a mean value of (1.57$\pm$0.01)$\times10^{-6}$ muons/(K s m$^2$). All
the fine structured deviations from the cosine have vanished within the
uncertainties.

\begin{table*}
  \caption{ \label{tab:modupar}
         List of parameters characterizing the annual modulation of the muon
         rate according to Ref.~\cite{borex_modu}. The theoretical value for the
         effective temperature coefficient for  \lngs\ is
         $\alpha_{T,\textrm{\lngs}}=0.92\pm0.02$.
}
  \begin{center}
  \begin{tabular}{lccccc}%\hline
  &&&&&\\[-3mm]\hline\hline
  &&&&&\\[-3mm]
    experiment
    &\textsc{Lvd}\cite{lvd_muons}&\textsc{Macro}\cite{macro_2003}&\textsc{Minos}\cite{minos_var}&\borexino\cite{borex_modu}
    &\gerda \\ 
&&&&&\\[-3mm] \hline
&&&&&\\[-3mm]
    site &\textsc{Lngs}-A&\textsc{Lngs}-B&Soudan&\textsc{Lngs}-C&
    \textsc{Lngs}-A\\
    duty cycle [yr]&8&7&5&4&2.5 \\
    published period& 2001-08&1991-97&2003-08&2007-11&2010-13 \\
    $E_{\textrm{thr}}$ [TeV] / [km.w.e.]&1.833 / 3.4&1.833 / 3.4&0.73/2.1&1.833
    /
    3.4&1.833 / 3.4 \\
    rate [10$^{-4}$/(s$\cdot$m$^2$)]&$3.31\pm0.03$& $3.22\pm0.08$&12.2374(3)~Hz
    &$3.41\pm0.01$&$3.47\pm0.07$ \\
    period [d]&$367\pm15$ &--&--&$366\pm3$& -- \\
    phase [d]&$185\pm15$ &--&--&$179\pm6$& $191\pm4$\\
    temp. data &Aer.Mil.&Aer.Mil.&\textsc{Ecwmf}&\textsc{Ecwmf}&
    \textsc{Ecwmf}/\textsc{Airs}
    \\
    $T_{\textrm{eff}}$ model contains&$\pi$&$\pi$&$\pi$+K&$\pi$+K& $\pi$+K\\
    correlation&0.53&0.91&0.90&0.62& 0.62/0.65\\
    $\alpha_T$&--&$0.91\pm0.07$&0.879$\pm$0.009 &$0.93\pm0.04$&$0.97\pm0.05$/\\
    &&&& &$0.93\pm0.05~~$~\\ \hline\hline
  \end{tabular}
\end{center}
\end{table*}%

\begin{figure}[t]
  \begin{center}
  \includegraphics[width=.49\textwidth]{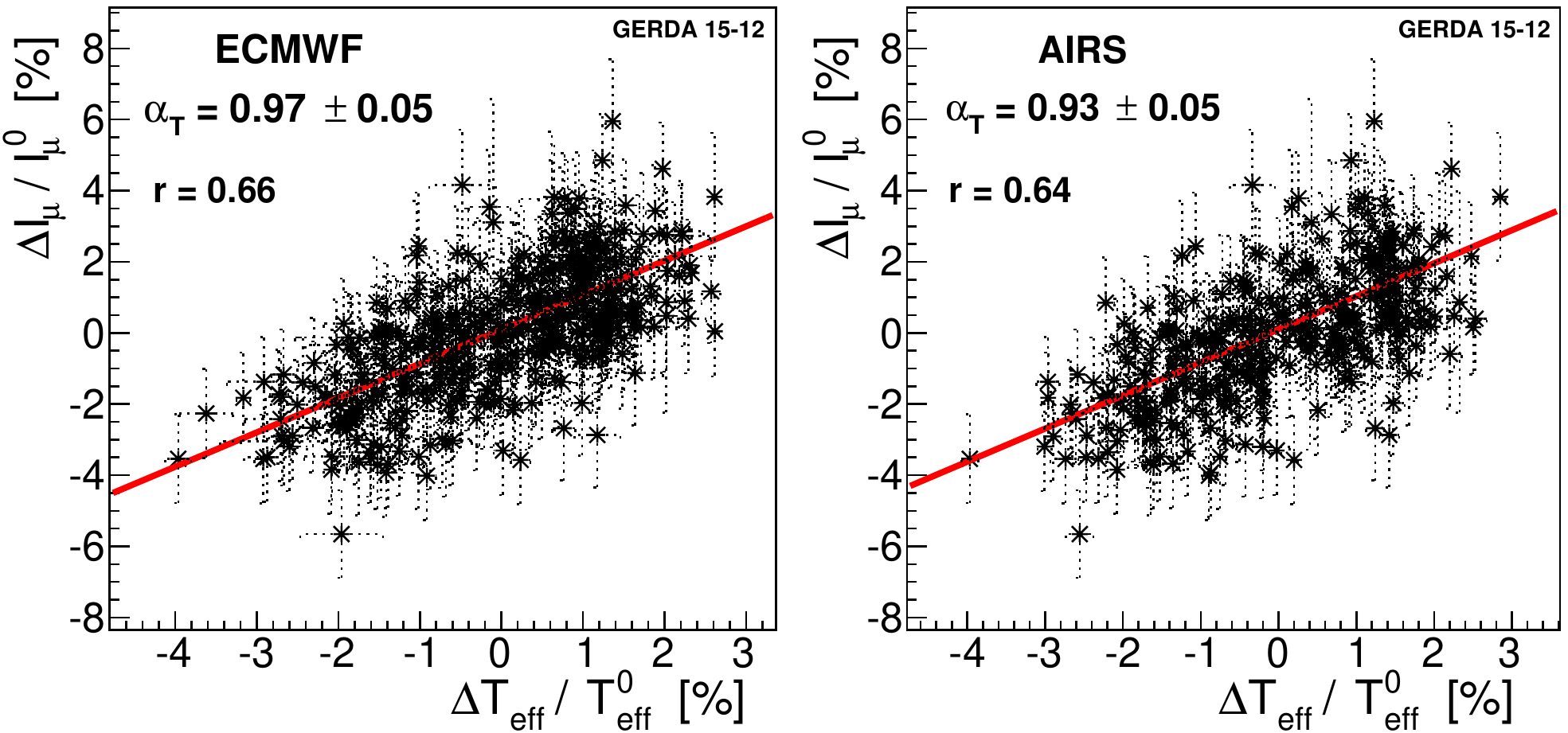}
  \caption{   \label{fig:alphas}
        Dependence of the change in muon rate on the change in effective
        temperature, for both sets of temperature data. A linear fit
        ($\chi^2$/ndf$_{\textrm{\textsc{Ecwmf}}}$=391/410,
        $\chi^2$/ndf$_{\textrm{\textsc{Airs}}}$=364/351) yields values for
        $\alpha_T$.
}
\end{center}
\end{figure}

The change in temperature versus the change in muon
flux can be quantified by the Pearson correlation coefficient $r$,
%\begin{equation}
% r=\frac{\sum_{i=0}^n (X_i -\bar{X})(Y_i
%    -\bar{Y})}{\sqrt{\sum_{i=0}^n (X_i-\bar{X})^2}\sqrt{\sum_{i=0}^n (Y_i
%      -\bar{Y})^2}},
%\end{equation}
which is $\sfrac{+}{-}1$ for a full positive/negative correlation and 0 for
uncorrelated values of $X={\Delta T_{\textrm{eff}}(t)}/{T_{\textrm{eff}}^0}$
and $Y={\Delta I_{\upmu}(t)}/{I_{\upmu}^0}$. A graphical representation can be
seen in Fig.~\ref{fig:alphas}, the coefficients $r$ of both data sets are
around 0.65 and thus a positive linear dependency exists. Therefore, the
change in temperature and muon flux can be written as:
\begin{equation}
  \frac{\Delta I_{\upmu}(t)}{I_{\upmu}^0}=\alpha_T\, \frac{\Delta
    T_{\textrm{eff}}(t)}{T_{\textrm{eff}}^0},
\end{equation}%
where $\alpha_T$ is an ``effective temperature coefficient''. Substituting
Eqs.~\ref{eq:sec} and~\ref{eq:four}, this coefficient becomes:
\begin{equation}
  \alpha_T = \frac{T_{\textrm{eff}}^0}{I_{\upmu}^0}\int_0^{\infty}dX\, W(X).
\end{equation}
allowing model predictions to access $\alpha_T$. Like $W(X)$, $\alpha_T$
depends on the threshold energy of the respective depth and on the amount of
muons from pion and kaon decay.  The values derived from linear fits to
the two data sets $\alpha_{T,\textrm{\textsc{Ecwmf}}}=0.97\pm0.05$ and
$\alpha_{T,\textrm{\textsc{Airs}}}=0.93\pm0.05$ are in agreement with each
other and with the values derived from
\borexino\ $\alpha_{T,bor}=0.93\pm0.04$~\cite{borex_modu} or \textsc{Macro}
$\alpha_{T,mac}=0.91\pm0.07$~\cite{macro_2003}.
%%%%%%%%%%%%%%%%%%%%%%%%%%%%%%%%%%%%%%%%%%%%%%%%%%%%%%%%%%%%%%%%%%%%%%%%%%%%%%%%
%%%%%%%%%%%%%%%%%%%%%%%%%%%%%%%%%%%%%%%%%%%%%%%%%%%%%%%%%%%%%%%%%%%%%%%%%%%%%%%%

This atmospheric model~\cite{grashorn,minos_var,kaon_pion2,kaon_pion}
containing both pion and kaon processes can be used to calculate a theoretical
value that amounts to $\alpha_{T,\textrm{\lngs}}=0.92\pm0.02$ for the
\textsc{Lngs} and that agrees well with both experimentally derived
values. The values derived from the present fit are summarized in
Tab.~\ref{tab:modupar} and are compared to the results of other experiments at
\textsc{Lngs} and Soudan which are in good agreement even though in some
analyses atmospheric models which only included muons produced by pion decay
are used.

\begin{figure}[t]
  \begin{center}
  \includegraphics[width=.49\textwidth]{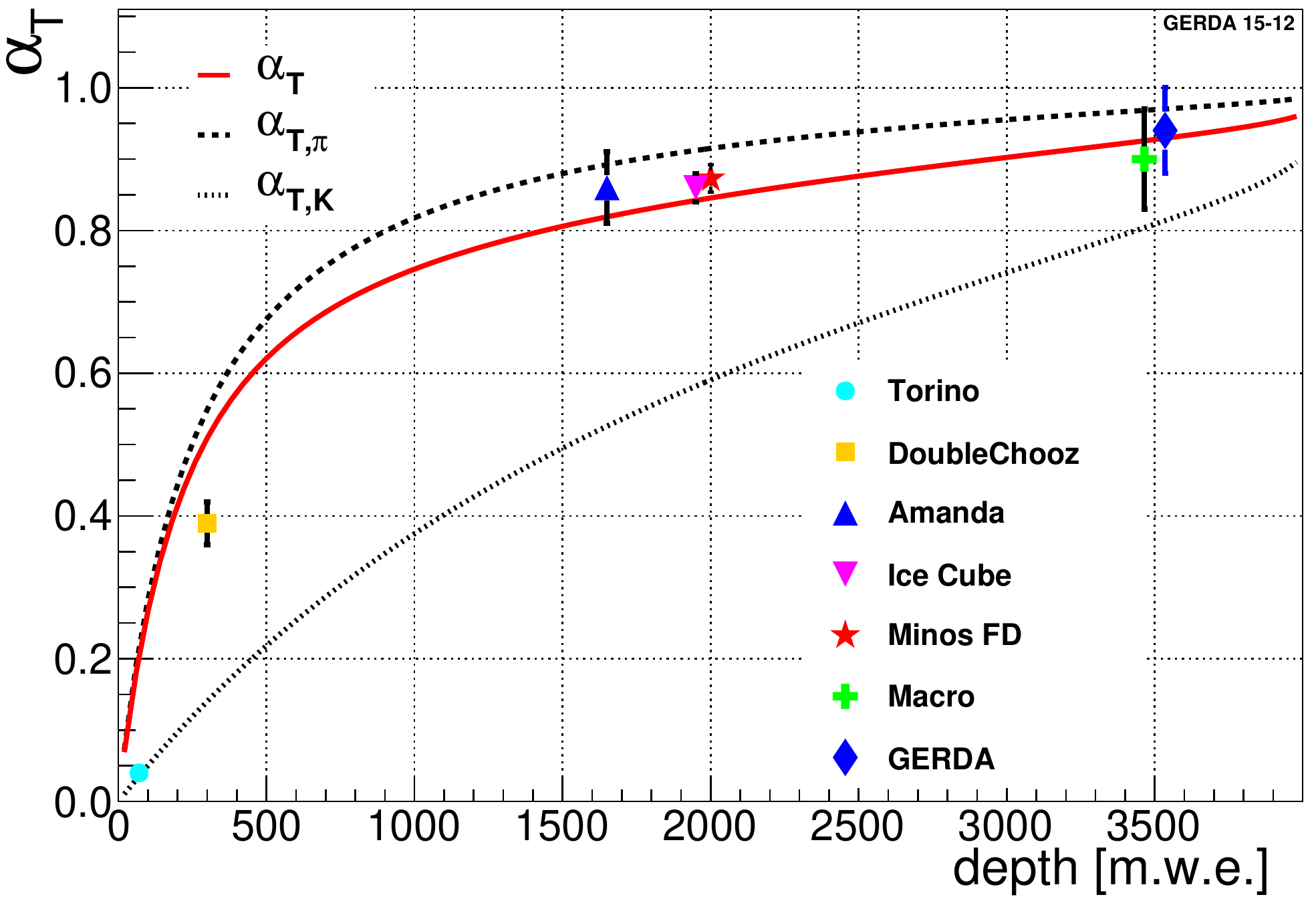}
  \caption{\label{fig:al_mwe}
      Correlation coefficient $\alpha_T$ as a function of depth. Experiments
      with different m.w.e. of rock overburden are listed such as
      Torino~\cite{torino_mod}, \textsc{Double Chooz}~\cite{dennis},
      \textsc{Amanda}~\cite{amanda_mod}, \textsc{IceCube}~\cite{icecube_mod},
      \textsc{Minos} far detector~\cite{minos_var},
      \textsc{Macro}~\cite{macro_2003} and \gerda\ (this work). \gerda\ and
      \textsc{macro} are located at the same depth but are drawn slightly
      apart for better visualization. The curves show muon generation models
      based on either purely pionic (dashed) or only kaonic (dotted)
      processes. The full red line notes the literature value for the
      atmospheric kaon/pion ratio~\cite{gaisser,kaon_pion}.
}
  \end{center}
\end{figure}

If the amount of rock overburden, i.e. the depth of the laboratory, is varied
in the atmospheric model, a relation between depth and $\alpha_T$ can be
calculated \cite{minos_var}. An additional factor in this calculation is the
ratio of pions to kaons produced in the atmosphere. Muons which originate from
kaons have a higher average energy and are thus less affected by the shielding
effect of the rock overburden. A graph of $\alpha_T$ as a function of depth of
observation (Fig.~\ref{fig:al_mwe}) allows for the extraction of the kaon to
pion ratio or a comparison of the measurements with the standard ratio. The
dotted lines in Fig.~\ref{fig:al_mwe} show the limits for pure kaon or pure
pion decays, i.e. $r_{K/\pi}=0$ or $\infty$. A model calculation with the
literature value for $r_{K/\pi}=0.149\pm0.06$~\cite{gaisser,kaon_pion} (red
line) describes all experiments below 500~m.w.e. well.
%%%%%%%%%%%%%%%%%%%%%%%%%%%%%%%%%%%%%%%%%%%%%%%%%%%%%%%%%%%%%%%%%%%%%%%%%%%%%%%%
\section{Summary}
The modulation of the muon flux in Hall A of \lngs\ was identified and
quantified using the muon veto data of the \gerda\ experiment during Phase~I and
before  for a total period of 806 live days.

%%%%%%%%%%%%%%%%%%%%%%%%%%%%%%%%%%%%%%%%%%%%%%%%%%%%%%%%%%%%%%%%%%%%%%%%%%%%%%%%
%%%%%%%%%%%%%%%%%%%%%%%%%%%%%%%%%%%%%%%%%%%%%%%%%%%%%%%%%%%%%%%%%%%%%%%%%%%%%%%%
In these data, two modulation effects with an overall influence on the muon
flux of 3--4\,\% could be identified: the additional muon flux caused by the
\textsc{Cngs} neutrino beam and the seasonal change in the muon rate caused by
temperature variation in the atmosphere which influences the muon production
mechanisms. A clear correlation between the \cngs\ $\nu_\mu$ beam and events
in the muon veto could be established thanks to the very precise GPS clock in
\textsc{Gerda}.  These coincident events were subtracted from the data set for
further analysis. Coincident \cngs -germanium events were found as well. Most
of these events were tagged by the muon veto. The number of untagged events is
consistent with the expected number of random coincidences.

An atmospheric model for the seasonal modulation of the muon flux due to
atmospheric changes was applied to the data. This model contains both pions
and kaons in the muon production mechanism. Two sets of climate data were used
to generate an effective temperature, which was found to be in direct relation
and in good correlation with the recorded muon flux variation. The results
were compared with other experiments and found to be in good agreement as
well. A mean muon rate of $I^0_{\upmu}= (3.477\pm0.002_{\textrm{stat}}
\pm0.067_{\textrm{sys}})\times10^{-4}$ /(s$\cdot$m$^{2}$) was found and the
correltation of the modulation with remperature was found to be
$\alpha_{T,\textrm{\textsc{Ecwmf}}} =0.97\pm0.05$ and
$\alpha_{T,\textrm{\textsc{airs}}} =0.93\pm0.05$ for the two data sets of
atmospheric data with $\alpha_{T,\textrm{\lngs}} =0.92\pm0.02$ being the
literature value for \textsc{Lngs}. The atmospheric modulation parameters were
compared to other experiments and agree well.

This data set provides a good basis for \gerda\ Phase~II, which aims at
10-fold lower background. The proven good timing will help to identify
cosmogenically produced isotopes. Additional analysis of muon data during
Phase~II of \gerda\ will allow for more sophisticated analyses including
systematic effects.

\section*{Acknowledgments}
The \gerda\ experiment is supported financially by the German Federal
Ministry for Education and Research (BMBF),
the German Research Foundation (DFG)
via the Excellence Cluster Universe, the Italian Istituto Nazionale
di Fisica Nucleare (INFN), the Max Planck Society (MPG),
the Polish National Science Center (NCN),
the Russian Foundation for Basic Research (RFBR),
and the Swiss National Science Foundation (SNF).
The institutions acknowledge also internal financial support.

The authors would like to thank  E.~Gschwendter and C.~Roderick for providing
the \textsc{Cngs} data set and the \textsc{Airs} and \textsc{Ecwmf} groups for
providing the climate data used in this work. The authors would also like to
thank D.~Dietrich and  M.~Wurm for many helpful discussions and valued input. 

%This work has been funded by the
%BMBF/PT-DESY grants 05CDVT1/8, 05A08VT1, 05A11VT3 and 05A14VT2.
%%%%%%%%%%%%%%%%%%%%%%%%%%%%%%%%%%%%%%%%%%%%%%%%%%%%%%%%%%%%%%%%%%%%%%%%%%%%%%%
%\section*{References}

%\bibliography{mybibfile}

\end{document}